\documentclass[journal=jpcbfk,manuscript=article,layout=twocolumn]{achemso}
\usepackage[version=3]{mhchem} 
\usepackage{graphicx}
\usepackage{amsfonts}
\usepackage{amssymb}
\usepackage{amsmath}
\usepackage{array}
\allowdisplaybreaks
\usepackage{textcomp}
\usepackage{longtable}
\usepackage{color}
\usepackage{psfrag}

\usepackage{epsfig}

\author{Duyu Chen}
\affiliation{Department of Chemistry, Princeton University, Princeton, New Jersey 08544, United States}
\alsoaffiliation{Physical Science in Oncology Center, Princeton University, Princeton, New Jersey 08544, United States}
\author{Yang Jiao}
\affiliation{Materials Science and Engineering, Arizona State University, Tempe, Arizona 85287, United States}
\author{Salvatore Torquato}
\email{torquato@princeton.edu}
\affiliation{Department of Chemistry, Princeton University, Princeton, New Jersey 08544, United States}
\alsoaffiliation{Physical Science in Oncology Center, Princeton University, Princeton, New Jersey 08544, United States}
\alsoaffiliation{Department of Physics, Princeton University, Princeton, New Jersey 08544, United States}
\alsoaffiliation{Program in Applied and Computational Mathematics, Princeton University, Princeton, New Jersey 08544, United States}
\alsoaffiliation{Princeton Institute of the Science and Technology of Materials, Princeton University, Princeton, New Jersey 08544, United States}

\title{Equilibrium Phase Behavior and Maximally Random Jammed State of Truncated Tetrahedra}

\abbreviations{Liquid-solid transition, Solid-solid transition, Adaptive-shrinking-cell scheme, Free-energy calculation, Prototypic glass}

\begin{document}

\begin{abstract}
Numerous recent investigations have been devoted to the
determination of the equilibrium phase behavior and packing
characteristics of hard nonspherical particles, including
ellipsoids, superballs, and polyhedra, to name but just a few
shapes. Systems of hard nonspherical particles exhibit a variety
of stable phases with different degrees of translational and
orientational order, including isotropic liquid, solid crystal,
rotator and a variety of liquid crystal phases. In this paper, we
employ a Monte Carlo implementation of the adaptive-shrinking-cell
(ASC) numerical scheme and free-energy calculations to ascertain
with high precision the equilibrium phase behavior of systems of
congruent Archimedean truncated tetrahedra over the entire range
of possible densities up to the maximal nearly space-filling
density. In particular, we find that the system undergoes two
first-order phase transitions as the density increases: first a
liquid-solid transition and then a solid-solid transition. The
isotropic liquid phase coexists with the Conway-Torquato (CT)
crystal phase at intermediate densities, verifying the result of a previous
qualitative study [\textit{J. Chem. Phys.}, {\bf 2011}, {\it 135}, 151101].
The freezing- and melting-point packing
fractions for this transition are respectively
$\phi_{F}=0.496\pm0.006$ and $\phi_{M}=0.591\pm0.005$. At higher
densities, we find that the CT phase undergoes another first-order
phase transition to one associated with the densest-known crystal,
with coexistence densities in the range $\phi \in [0.780 \pm
0.002, 0.802 \pm 0.003]$. We find no evidence for stable
rotator (or plastic) or nematic phases. We also generate the
maximally random jammed (MRJ) packings of truncated tetrahedra,
which may be regarded to be the glassy end state of a rapid
compression of the liquid. Specifically, we systematically study
the structural characteristics of the MRJ packings, including the
centroidal pair correlation function, structure factor and orientational pair
correlation function. We find that
such MRJ packings are hyperuniform with an average
packing fraction of 0.770, which is considerably larger
than the corresponding value for identical spheres ($\approx 0.64$). We conclude with some simple observations concerning what types of phase transitions might be expected in general hard-particle systems based on the
particle shape and which would be good glass formers.

\end{abstract}


\maketitle

\section{I. INTRODUCTION}
Hard-particle systems have served as useful models of
low-temperature states of matter, including liquids, \cite{frenkel1997understanding} crystals, \cite{chaikin2000principles} glasses, \cite{chaikin2000principles, zallen1983} granular media, \cite{torquato2002random, edwards1994granular} heterogenous materials, \cite{torquato2002random} and powders. \cite{powder} Understanding the equilibrium and nonequilibrium properties of hard particle systems is of great interest. This is reflected by the numerous studies devoted to
these topics for nonspherical particles that span a wide range of
shapes, including ellipsoids, superballs, and polyhedra.
\cite{frenkel1997understanding, bautista2013further,
frenkel1985phase, ni2012phase, gantapara2013phase,
jiao2011communication, vega2008determination, sal_2009nature, torquato2009dense, mrj1, mrj2, mrj3, escobedo1, nonspherical1} Nanoparticles and colloidal particles of
various shapes can now be routinely synthesized in the laboratory.
\cite{xia2012symmetry, rossi2011cubic, quan2010superlattices,
schweiger2005self}

In general, a packing is defined as a large collection of
nonoverlapping solid objects (particles) in d-dimensional
Euclidean space $\mathbb{R}^{d}$. The associated packing fraction
(or density) $\phi$ is defined as the fraction of space
$\mathbb{R}^{d}$ covered by the particles. The densest packing of
a specific particle shape, which is usually achieved by an ordered
arrangement depending on the particle symmetry,
\cite{sal_2009nature, torquato2009dense} is the thermodynamic stable phase of that
shape in the infinite-pressure limit, \cite{torquato2010jammed}
and thus provides the starting point to determine the entire
high-density phase behavior of the system. On the other hand, the
maximally random jammed (MRJ) state of packing can be considered
to be a prototypical glass. \cite{torquato2000mrj, torquato2010jammed} Roughly
speaking, MRJ packings can be obtained by compressing liquid
configurations at the largest possible rate such that the packing
is strictly jammed (i.e., mechanically stable).
\cite{torquato2010jammed}

\begin{figure}[htp]
\includegraphics[width=0.45\textwidth]{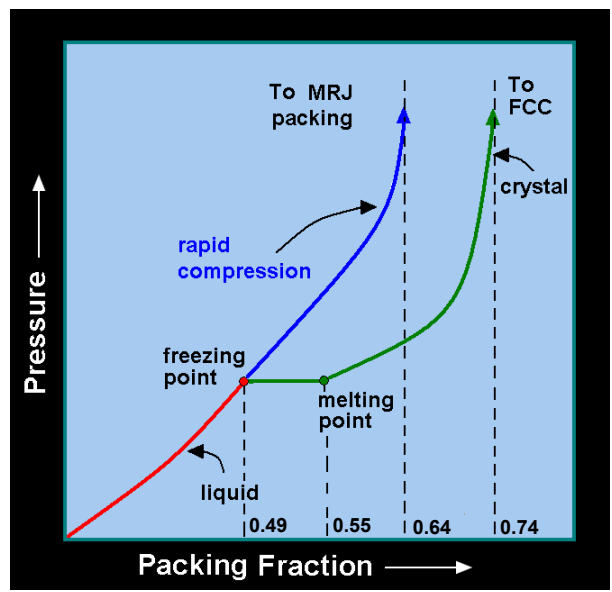}
\caption{Schematic plot for the pressure of the
hard-sphere system as a function of packing fraction along the stable liquid and crystal branches,
and along a metastable branch ending at the MRJ state in the infinite-volume limit. This figure is adapted from ref~\citenum{torquato2010jammed}.}
\label{fig_sphere_phase}
\end{figure}

The hard-sphere model in $\mathbb{R}^3$ has a venerable history.
\cite{frenkel1997understanding, torquato2002random, alder_phase, hanson_liquid, frenkel1987new, Ri96, speedy1998pressure, mau_huse,torquato2010jammed,jadrich1, jadrich2}
It is well known from numerical simulations that such a system
exhibits a first-order liquid-crystal phase transition as the
density increases along the liquid branch. The associated
freezing- and melting-point packing fractions have been
determined to be around 0.49 and 0.55, respectively, by both
pressure and free-energy calculations.
\cite{frenkel1997understanding,torquato2010jammed} At the maximal
density, free-energy calculations have been used to demonstrate
that the face-centered cubic crystal (FCC) is very slightly more stable
than the hexagonal close-packed crystal. \cite{frenkel1987new,
speedy1998pressure, mau_huse} Upon rapid compression from a liquid
configuration, the system falls out of equilibrium and follows a
metastable branch, whose end state is presumably the MRJ state in the infinite-volume limit.
\cite{Ri96, torquato2010jammed} The aforementioned
equilibrium and nonequilibrium properties of the hard-sphere
system are schematically illustrated in Figure \ref{fig_sphere_phase}.

Since nonspherical particles have both translational and
rotational degrees of freedom, they usually have a richer phase
diagram than spheres, i.e., the former can possess different degrees
of translational and orientational order. For example, a rotator (or plastic) phase is
one in which particles possess translational order but can rotate
freely. \cite{vega1998solid, bolhuis1997entropy, marechal2010phase}
A nematic phase is one in which the particles are aligned (i.e.,
with orientational order) while the system lacks any long-range
translational order. \cite{onsager1949effects,rego2009asymmetric}
A smectic phase is one in which particles have ordered orientations and possess
translational order in one direction. \cite{smectic_ref}
The types of phases formed by hard nonspherical particles are
influenced by many factors. It is well known that
entropy plays a principal role in determining the phase behavior of hard-particle systems. For example, spheroids (ellipsoids of revolution) with needle-like shapes exhibit a liquid-nematic phase transition at low
densities. \cite{onsager1949effects} While more recent studies
\cite{bautista2013further,frenkel1985phase} have also revealed
that spheroids with shapes closer to spheres
exhibit a liquid-rotator crystal phase transition at intermediate
densities. It has been demonstrated that the local curvature of the particle shape
contributes to the stabilization of rotator phases by studies of superballs
at intermediate densities. \cite{ni2012phase, superball2} Gantapara et al.
\cite{gantapara2013phase} showed that the phase diagram of
truncated cubes exhibits a rich diversity in crystal structures
that depend sensitively on the amount of truncation. When the interactions
are dominated by hard-particle-like repulsions, such as in certain
polymer systems, the role of entropy is significant as well. \cite{polymer}

\begin{table}
\caption{Some Geometrical Properties of the Archimedean Truncated
Tetrahedron with Side Length $a$.}
\begin{center}
\begin{tabular}{>{\centering\arraybackslash}m{3cm}>{\centering\arraybackslash}m{4cm}} \\ \hline\hline
Inradius, $r_{in}$ & $\frac{\sqrt{6}a}{4}$ \\
\hline
Circumradius, $r_{out}$ & $\frac{\sqrt{22}a}{4}$\\
\hline
Asphericity, $\gamma$ & $\frac{\sqrt{33}}{3}$ \\
\hline
Radius of mean curvature, $\bar{R}$ & $\frac{9}{4\pi}\cos^{-1}(\frac{1}{3})a$\\
\hline
Surface Area, $S$ & $7\sqrt{3}a^2$ \\
\hline
Volume, v & $\frac{23}{12}\sqrt{2}a^3$\\
\hline
Scaled exclusion volume, $\textnormal{v}_{ex}/\textnormal{v}$     & $2+\frac{378}{23\pi}\sqrt{\frac{3}{2}}\cos^{-1}(\frac{1}{3})$ \\
\hline\hline
\end{tabular}
\end{center}
\label{tab_geo_param}
\end{table}

One aim of this paper is to determine systematically the
equilibrium phase behavior of the Archimedean truncated
tetrahedra for the entire density range. An Archimedean truncated tetrahedron (henceforth often
called a truncated tetrahedron for simplicity) is obtained by
truncating the corners of a regular tetrahedron with edge length
that is one-third of the edge length of the original tetrahedron, and it therefore
has four regular triangular faces and four hexagonal faces. It is
of particular interest because it is the only nonchiral
Archimedean solid without central symmetry and, as we will
discuss, packings of truncated tetrahedra can nearly fill all of
space. Some geometrical properties of the truncated tetrahedron,
including the inradius $r_{in}$, circumradius $r_{out}$, asphericity $\gamma \equiv r_{out}/r_{in}$,
radius of mean curvature $\bar{R}$, \cite{sal_perco1} surface area $S$, volume v, and scaled
exclusion volume $\textnormal{v}_{ex}/\textnormal{v}$, \cite{sal_perco1} are summarized in Table \ref{tab_geo_param}.

\begin{figure}[!ht]
\begin{center}
$\begin{array}{c@{\hspace{0.5cm}}c}\\
\includegraphics[width=0.25\textwidth]{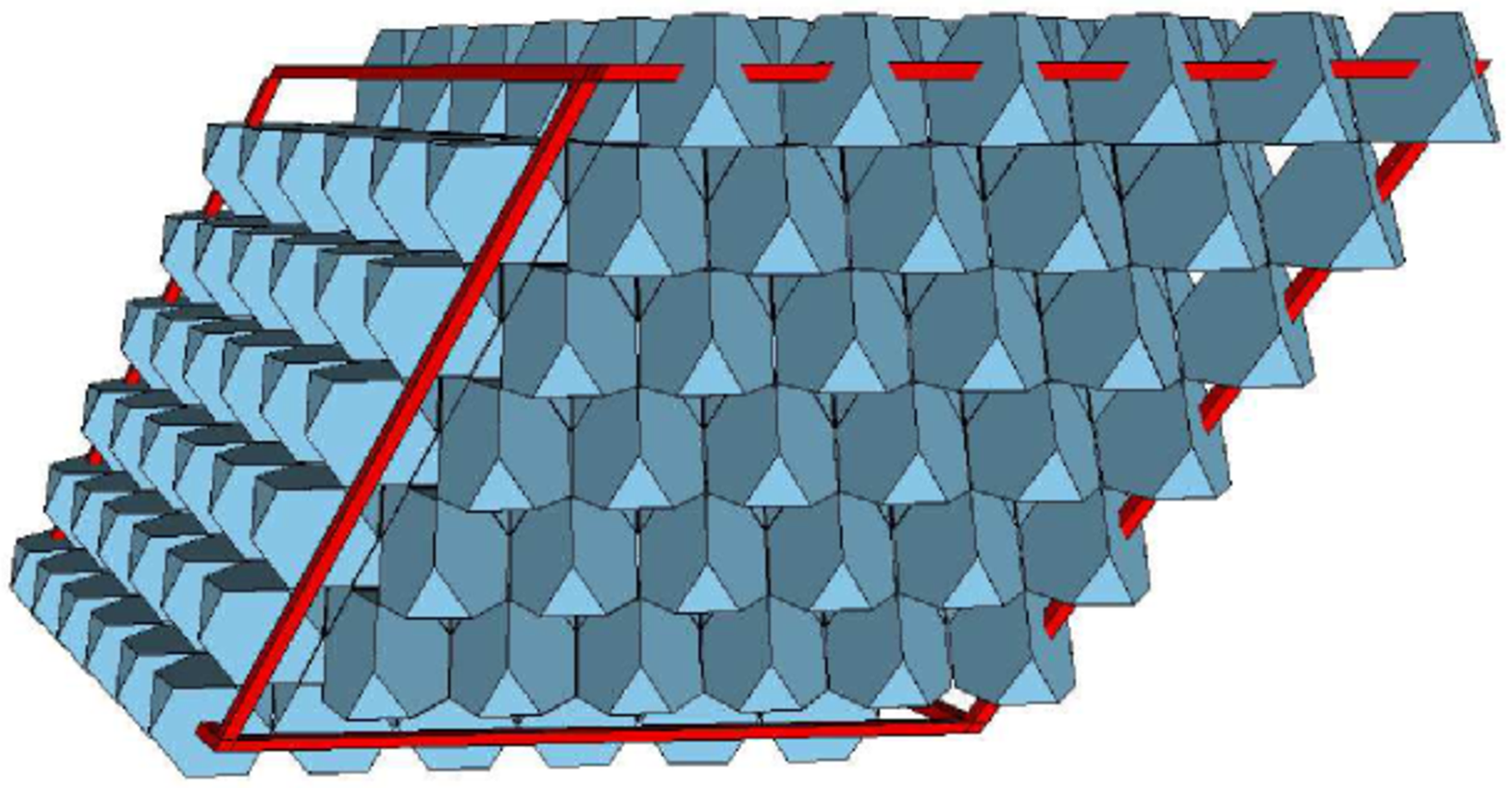} &
\includegraphics[width=0.17\textwidth]{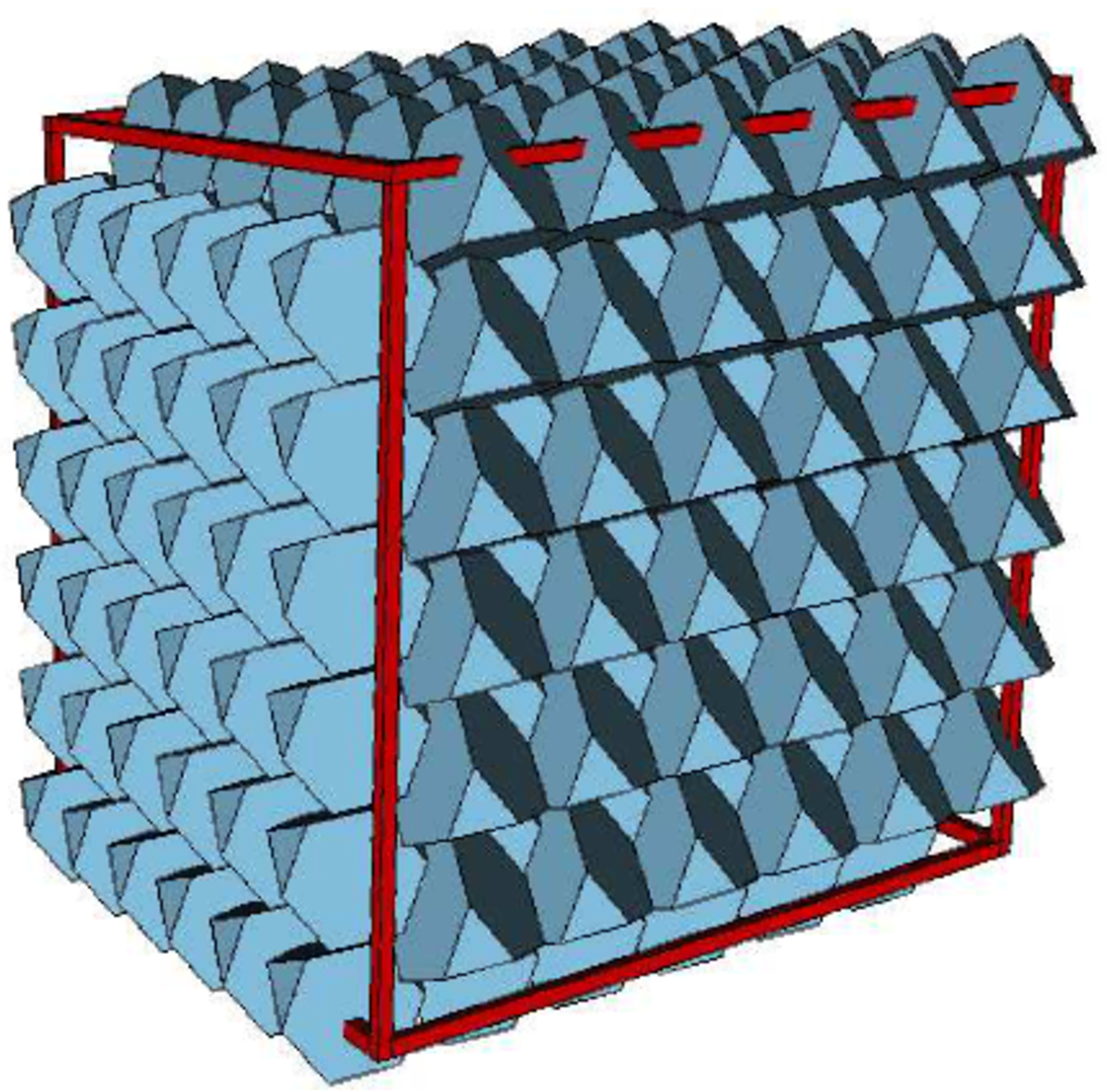} \\
\mbox{\bf (a)} & \mbox{\bf (b)}
\end{array}$
\end{center}
\caption{(Color online) (a) Portion of the CT crystal. (b) Portion of the densest known crystal.} \label{fig_phases}
\end{figure}

Jiao and Torquato \cite{jiao2011communication} have recently
analytically constructed the densest known packing of such
polyhedra with $\phi = 207/208 = 0.995192...$. Given that this
packing fraction is nearly equal to unity and truncated tetrahedra
cannot tile three-dimensional Euclidean space $\mathbb{R}^{3}$,
this densest known packing is likely the densest packing for such solids. In the same paper, the
equilibrium melting properties of truncated tetrahedra were also
qualitatively studied by simply monitoring the structural changes
upon decompression from the densest known packing. They found that
the system apparently undergoes a transition to another crystal phase, namely
the one associated with the packing of truncated tetrahedra discovered
by Conway and Torquato \cite{conway2006packing} (henceforth
referred to as ``CT crystal'') as well. Figure \ref{fig_phases}
shows a portion of the CT packing and the densest known packing.
However, only free-energy calculations can yield quantitatively
accurate information about the phase behavior of truncated
tetrahedra over the entire range of densities.

In the present work, we employ a Monte Carlo implementation of the
adaptive-shrinking-cell (ASC) numerical scheme
\cite{sal_2009nature,torquato2009dense} and free-energy calculations
\cite{frenkel1997understanding} to investigate with high precision
the equilibrium phase behavior of truncated tetrahedra and to
verify whether the types of phases found in ref~\citenum{jiao2011communication} are indeed thermodynamically
stable. Consistent with the findings of ref~\citenum{jiao2011communication}, we find that the system
undergoes two first-order phase transitions as the density
increases: first a liquid-solid transition and then a solid-solid
transition. The isotropic liquid phase coexists with the
Conway-Torquato (CT) crystal phase with densities in the range
$\phi \in [0.496\pm0.006, 0.591\pm0.005]$. At higher densities,
the CT phase coexists with the one associated with the
densest-known crystal with densities in the range $\phi \in [0.780
\pm 0.002, 0.802 \pm 0.003]$. We find no evidence for stable
rotator or nematic phases. Moreover, we also generate
maximally random jammed packings of truncated tetrahedra using ASC
simulations with a sufficient large compression rate and study
their characteristics, including the packing fraction, centroidal pair
correlation function, structure factor and orientational correlation
function. We find that the MRJ
packings are hyperuniform \cite{hyperuniform} (see section V for definitions and
details). In closing, we explain why truncated tetrahedra are expected to have the two types of first-order phase transitions reported here.


The rest of the paper is organized as follows: In section II, we
discuss the simulation methods used in studying the phase
behavior of truncated tetrahedra. In section III, we
discuss the structural descriptors used to characterize equilibrium and
MRJ packings of truncated tetrahedra. In section IV, we study the
equilibrium phase behavior of truncated tetrahedra using ASC
simulations and free-energy calculations. In section V, we generate
and characterize MRJ packings of truncated tetrahedra. In section VI,
we provide concluding remarks and make some simple observations concerning what types of phase transitions
might be expected in general hard-particle systems based on the particle shape.

\section{II. SIMULATION PROCEDURES FOR PHASE BEHAVIOR}

\subsection{A. Adaptive-Shrinking-Cell Monte Carlo Method}

The adaptive-shrinking-cell (ASC) Monte Carlo method is employed
to equilibrate hard truncated tetrahedra at different packing
fractions. While we sketch the procedure here, the reader
is referred to ref~\citenum{torquato2009dense} for additional details.
Along the liquid branch, initially a system at each packing fraction
is generated by compressing dilute disordered particle configurations with $\phi<0.1$
in a simulation box subject to periodic boundary conditions.
Along the crystal branch, initially a system at each packing fraction
is generated by dilating the densest crystal of the particles in a
simulation box subject to periodic boundary conditions. Our ASC scheme is capable
of shrinking the simulation box, but we do not employ that feature for the determination of the phase behavior (just use it for the generation of MRJ states, as described in section V). The initial systems
at fixed densities are equilibrated by particle trial
moves and volume-preserving shear deformations of the simulation box.
Specifically, for a fixed simulation box, each particle is sequentially randomly displaced
and rotated by small amounts. A trial move is rejected if it results in overlap between a pair of
particles and is accepted otherwise. The \textit{separation axis
theorem} (SAT) \cite{golshtein1996modified} is used to check overlaps.
After a prescribed number of trial-move cycles, the boundary
of the periodic simulation box (fundamental cell) is allowed to
deform by specified small amount. Such a boundary move is
accepted if no overlaps between any pair of particles in the
system occur and is rejected otherwise. The boundary
deformation is represented by a symmetric strain tensor, whose
trace (i.e., the sum of the diagonal components) corresponds to
the volume change (i.e., compression or expansion) of the
fundamental cell, and the off-diagonal components correspond to
the shape change (i.e., deformation) of the fundamental cell.
The equilibrium pressure is obtained as discussed in the ensuing
section.

\subsection{B. Pressure Calculation}

The equilibrium pressure of the truncated-tetrahedron system in
our NVT simulation is computed from the distribution of
interparticle gaps. Following ref~\citenum{eppenga1984monte},
the scaled pressure $Z$ is given by
\begin{equation}
\label{eq_pressure} Z \equiv \frac{p}{\rho k_{B}T} =
1+\frac{\phi\alpha}{2}
\end{equation}
where $\rho = N/V$ is the number density of the system, $T$ is the temperature,
and $k_{B}$ is the Boltzmann's constant. The parameter $\alpha$ depends on the
particle shape and is computed from the following relation:
\begin{equation}
\ln P_{1}(\triangle\phi)=\ln\alpha-\alpha\triangle\phi
\end{equation}
where $P_{1}(\triangle\phi)$ is the probability that a given
particle first overlaps with another particle if the packing
fraction of the system is increased by $\triangle\phi$, regardless
of other overlapping pairs of particles.

In practice, one just needs to compute for each particle the
minimal compression (i.e., change of volume) leading to overlap
between it and its nearest neighbor. This process is
repeated for every particle in the system and a histogram of the
distribution of minimal interparticle gaps is then obtained.

\subsection{C. Free Energy Calculations}

\subsubsection{\textit{Free Energy Calculations of the Liquid Phase}}

To compute the free energy of the liquid phase, we first use
pressure calculations discussed in section II.B to obtain the
associated dimensionless equation of state (EOS) for the scaled pressure,
i.e., $Z=p/(\rho k_{B}T)$. Then we integrate the EOS from
a low-density reference state with packing fraction $\phi_{0}$ to a specific $\phi$
to get the associated Helmholtz free energy,
\cite{vega2008determination} i.e.,
\begin{equation}
\label{eq_energy_liquid}
\frac{A(\phi)}{Nk_{B}T}=\frac{A(\phi_{0})}
{Nk_{B}T}+\int_{\phi_{0}}^{\phi}\frac{p(\phi')\textnormal{v}}{k_{B}T{\phi'}^{2}}\mathrm{d}\phi'
\end{equation}
where
\begin{equation}
A(\phi_{0})/(Nk_{B}T)=\mu(\phi_{0})/(k_{B}T)-p(\phi_{0})\textnormal{v}/(k_{B}T\phi_{0})
\end{equation}
and v is the volume of a particle, $\mu(\phi_{0})$ the chemical
potential at packing fraction $\phi_{0}$, which is calculated by
the Widom's particle insertion method. \cite{widom1963some} In
this paper, we use a reference system at $\phi_{0} = 0.10$ for
liquid-phase free energy calculations.

\subsubsection{\textit{Free Energy Calculations of a Solid Phase}}

To compute the free energy of a solid phase, we employ the
standard NVT Einstein crystal method. \cite{vega2008determination,
ni2012phase, frenkel1997understanding, noya2008computing}
Specifically, we construct a reversible path between the actual
hard-particle crystal and an ideal Einstein crystal, which allows
us to calculate the free-energy difference between these two
systems. Since the free energy of the ideal Einstein crystal is
known analytically, we can thus obtain the absolute value of the
free energy for the hard-particle crystal, i.e.,
\begin{equation}
\label{eq_energy_solid}
\frac{A(N,V,T)}{k_{B}T}=\frac{A_{E}(N,V,T)}{k_{B}T}-\int_{0}^{\lambda_{max}}\mathrm{d}\lambda
\langle\frac{\partial U_{E}(\lambda)}{\partial\lambda}\rangle
\end{equation}
where $\langle ...\rangle$ denotes ensemble average of systems
with coupling potential $U_{E}$, and $\lambda_{max}$ is the
maximum coupling constant that is sufficiently strong to suppress particle collisions.
When $\lambda=\lambda_{max}$, the system behaves like an ideal Einstein crystal.
When $\lambda=0$, there is no coupling potential and the system behaves as the real crystal
under consideration. The free energy of the ideal Einstein crystal $A_{E}(N,V,T)$ is given
by
\begin{equation}
\begin{split}
\label{eq_energy_einst}
\frac{A_{E}(N,V,T)}{k_{B}T}=-\frac{3(N-1)}{2}\ln(\frac{\pi k_{B}T}{\lambda_{max}})+N\ln(\frac{\Lambda_{t}^{3}\Lambda_{r}}{\sigma^{4}})\\ +\ln(\frac{\sigma^{3}}{VN^{1/2}})-N\ln\{\frac{1}{8\pi^2}\int\mathrm{d}\theta\mathrm{d}\phi\mathrm{d}\chi \\
\sin\theta\exp[-\frac{\lambda_{max}}{k_{B}T}(\sin^2\psi_{\bf{a}}+\sin^2\psi_{\bf{b}})]\}
\end{split}
\end{equation}
where $\Lambda_{t}$ and $\Lambda_{r}$ are the translational and
rotational thermal de Broglie wavelengths, respectively, and both
are set to 1 in our simulations; $\sigma=1/\textnormal{v}^{1/3}$ is the
characteristic length of the particle with volume v; $\theta$,
$\phi$ and $\chi$ are the Euler angles \cite{gray1984theory}
defining the orientation of the particle with respect to the
reference orientation in the reference lattice; $\psi_{\bf{a}}$
and $\psi_{\bf{b}}$ are the minimal angles formed by the two
reference vectors $\bf{a}$, $\bf{b}$ and the characteristic
vectors of a particle defining its orientation. The potential
$U_{E}(\lambda)$ characterizes the coupling between the hard
particles to their reference lattice sites and reference
orientations, \cite{noya2010stability,ni2012phase} i.e.,
\begin{equation}
\label{eq_coupling_field} U_{E}(\lambda)=\lambda
\sum\limits_{i=1}^N
[({\bf{r}}_{i}-{\bf{r}}_{i0})^2/\sigma^2+(\sin^2\psi_{i\bf{a}}+\sin^2\psi_{i\bf{b}})]
\end{equation}
where ${\bf{r}}_{i}-{\bf{r}}_{i0}$ is the displacement of particle $i$ with respect to its reference lattice site.



\subsubsection{\textit{Validation of Free Energy Calculation}}

\begin{figure}[!ht]
\begin{center}
$\begin{array}{c}
\includegraphics[width=0.35\textwidth]{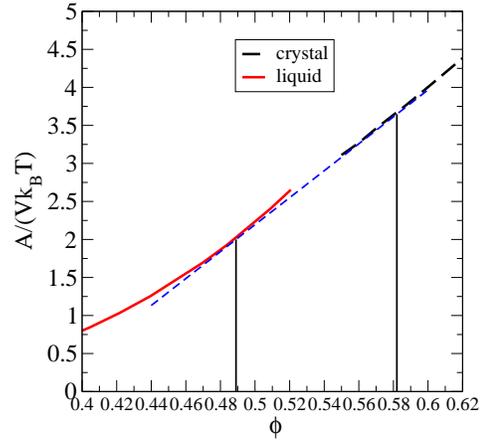}
\end{array}$
\end{center}
\caption{(Color online) Reduced Helmholtz free energy per unit volume $A/(Vk_{B}T)$ as a function of packing fraction
$\phi$ for hard octahedra. The liquid phase is found to coexist with the Minkowski crystal. The
freezing- and melting-point packing fractions are estimated
to be $0.489\pm0.004$ and $0.582\pm0.008$, respectively.}
\label{fig_octa_energy}
\end{figure}

To validate our implementation of the free energy calculations, we
have calculated the free energy of hard regular octahedra within intermediate
density ranges, where it has been previously shown that there
is a first-order liquid to Minkowski crystal \cite{octa_densest}
phase transition.\cite{gantapara2013phase,ni2012phase} Our results are
shown in Figure \ref{fig_octa_energy}. On the basis of our free energy
calculations, we estimate the freezing- and melting-point packing fractions of
hard octahedra to be $0.489\pm0.004$ and $0.582\pm0.008$,
respectively, which agree well with previously reported values in
the literature. \cite{gantapara2013phase,ni2012phase}

\section{III. STRUCTURAL DESCRIPTORS}

In this section, we present a number of structural
descriptors that quantitatively characterize various packing
structures of truncated tetrahedra that are associated with the
equilibrium and nonequilibrium phases of the system. In addition,
we briefly review the ``hyperuniformity'' concept and its
quantification, which appears to be a universal structural
characteristic of general maximally random jammed packings.

\subsection{A. Orientational Pair Correlation Function}
The emergence of orientational order in a MC simulations is usually a strong
indication of a possible first-order phase transition.
To quantify the orientational order in the system of
truncated tetrahedra at different densities, we use an
orientational pair correlation function, which we define as:
\begin{equation}
\label{eq_orient_corr} g_{orient}(r)=\frac{\sum\limits_{i=1}^N
\sum\limits_{j\not=i}^N
\delta(r-r_{ij})(\sin^2\alpha_{ij}^{F_{1}}+\sin^2\alpha_{ij}^{F_{2}})/2}{\sum\limits_{i=1}^N
\sum\limits_{j\not=i}^N \delta(r-r_{ij})}
\end{equation}
where $\alpha_{ij}^{F_{1}}$ is the minimum angle formed by any of
the characteristic vectors associated with truncated tetrahedron $i$ and
any of the characteristic vectors associated with truncated tetrahedron $j$, and
$\alpha_{ij}^{F_{2}}$ is the second minimum besides $\alpha_{ij}^{F_{1}}$
($\alpha_{ij}^{F_{1}}\leq\alpha_{ij}^{F_{2}}$).
We define a characteristic vector associated with a truncated tetrahedron
as the unit vector pointing from the particle centroid to the center of one of the four hexagonal faces
of a truncated tetrahedron. The superscript $F_{1}$ denotes such pair of characteristic vectors passing through
the hexagonal face centers that minimizes the angle between them, and $F_{2}$ denotes the next pair that minimizes
the angle besides the pair associated with $F_{1}$. The orientational
pair correlation function will be employed to suggest possible stable
phases for subsequent free energy calculations.

\subsection{B. Structure Factor and Centroidal Pair Correlation Function}

The ensemble-averaged structure factor of infinite point
configurations in $d$-dimensional Euclidean space at number
density $\rho$ is defined via
\begin{equation}
\label{eq_Sk} S({\bf k})= 1+ \rho {\tilde h}({\bf k})
\end{equation}
where ${\tilde h}({\bf k})$ is the Fourier transform of the total
correlation function $h({\bf r}) = g_2({\bf r})-1$ and $g_2({\bf
r})$ is the centroidal pair correlation function of the system. Note that
definition \ref{eq_Sk} implies  that the forward scattering
contribution to the diffraction pattern is omitted. For
statistically homogeneous and isotropic systems, the focus of this
paper, $g_2$ depends on the radial distance $r \equiv |{\bf r}|$
between the points as well as the number density
$\rho$.

For computational purposes, the structure factor $S({\bf k})$ for
a given finite point configuration can be obtained directly from
the positions of the points ${\bf r}_j$, \cite{chase} i.e.,
\begin{equation}
S({\bf k}) = \frac{1}{N} \left |{\sum_{j=1}^N \exp(i {\bf k} \cdot
{\bf r}_j)}\right |^2 \quad ({\bf k} \neq {\bf 0}) \label{coll}
\end{equation}
where $N$ is the total number points in the system (under periodic
boundary conditions) and ${\bf k}$ is the wave vector. Note that
the forward scattering contribution (${\bf k} = 0$) in equation
\ref{coll} is omitted, which makes relation \ref{coll}
completely consistent with the definition \ref{eq_Sk} in the
ergodic infinite-system limit. For statistically homogeneous and
isotropic systems, the focus of this paper, the structure factor
$S(k)$ only depends on the magnitude of the scalar wavenumber $k =
|{\bf k}| = 2\pi n/L$, where $n = 0, 1, 2\ldots$ and $L$ is the
linear size of the system.

\subsection{C. Hyperuniform Systems}

The small-$k$ behavior of the structure factor $S(k)$ encodes
information about large-scale spatial correlations in the system
and in the limit $k \rightarrow 0$ determines whether the system
is hyperuniform. Specifically, an infinite point configuration in
$d$-dimensional Euclidean space is {\it hyperuniform} if
\begin{equation}
\label{eq_Sk2} \lim_{k\rightarrow 0}S(k) = 0
\end{equation}
which implies that the infinite-wavelength density fluctuations of
the system (when appropriately scaled) vanish. \cite{hyperuniform}

A hyperuniform point configuration has the property that the
variance in the number of points in an observation window $\Omega$
grows more slowly than the volume of that window. \cite{hyperuniform} In
the case of a spherical observation window of radius $R$, this
definition implies that the local number variance $\sigma^2(R)$
grows more slowly than $R^d$ in $d$ dimensions. The local number variance
of a statistically homogeneous point
configuration in a spherical observation window is given exactly
by
\begin{equation}\label{numvar}
\sigma^2(R) = \rho \textnormal{v}(R) \left[1+\rho \int_{\mathbb{R}^d}
h(\mathbf{r}) \alpha(\mathbf{r}; R) d\mathbf{r}\right]
\end{equation}
where v$(R)$ is the volume of a spherical window of radius $R$ and
$\alpha(\mathbf{r}; R)$ is the \emph{scaled intersection volume}, i.e., the
intersection volume of two spheres of radius $R$ separated by a
distance $r$ divided by the volume of a sphere v$(R)$.

It has been shown that the number variance in eq \ref{numvar}, under
certain conditions, admits the following asymptotic scaling:
\cite{hyperuniform}
\begin{equation}
\label{numasymp}
\begin{split}
\sigma^2(R) = 2^d \phi\{A(\frac{R}{D})^d + 
B(\frac{R}{D})^{d-1} + \\
 o[(\frac{R}{D})^{d-1}]\}
\end{split}
\end{equation}
where
\begin{align}
A &= 1+\rho\int_{\mathbb{R}^d} h(\mathbf{r}) d\mathbf{r} =
\lim_{\lVert\mathbf{k}\rVert\rightarrow 0}
S(\mathbf{k})\label{An}
\end{align}
$D$ is a characteristic microscopic length associated with the
point configuration (e.g., the average nearest-neighbor distance
between the points) and $o(x)$ denotes all terms of order less
than $x$. Clearly, when the coefficient $A = 0$, i.e., $\lim_{{\bf
k}\rightarrow 0}S({\bf k}) = 0$ satisfies the requirements for
hyperuniformity. The relation in \ref{An} then implies that
hyperuniform point patterns do not possess infinite-wavelength
density fluctuations (when appropriately scaled) and hence from
\ref{numasymp} the number variance scales as the surface area of
the window for large $R$, i.e., $\sigma^2(R) \sim R^{d-1}$ in the
large-$R$ limit. Equations 13 and 14 are valid for all periodic point
patterns (including perfect lattices), quasicrystals, and disordered
systems in which the pair correlation function $g_2$ decays to
unity exponentially fast. \cite{hyperuniform} The degree to which
large-scale density fluctuations are suppressed enables one to
rank order crystals, quasicrystals and special disordered systems.
\cite{hyperuniform, chase}

Disordered hyperuniform structures can be
regarded as new states of disordered matter in that they behave
more like crystals or quasicrystals in the manner in which they
suppress density fluctuations on large length scales, and yet are
also like liquids and glasses in that they are statistically
isotropic structures with no Bragg peaks. Thus, disordered hyperuniform
materials can be regarded to possess a ``hidden order''
that is not apparent on short length scales and are endowed with novel physical properties. Such states of matter can be arrived at via both equilibrium and nonequilibrium routes and include fermionic
ground states, \cite{hyperuniform1} classical disordered ground states, \cite{hyperuniform2} and  MRJ particle packings. \cite{mrj3, chase_prl, chase2, chase3, chase4} Disordered hyperuniform dielectric network materials have been shown to possess large and complete photonic band gaps. \cite{hyperuniform3} More recently, it has been demonstrated that this exotic state of matter arises in the  photoreceptor patterns in avian retinas. \cite{hyperuniform4}

For disordered hyperuniform systems with a total correlation
function $h(r)$ that does not decay to zero exponentially fast,
other dependencies of the number variance on $R$ may be observed.
More generally, for any reciprocal power law,
\begin{equation}
S(k) \sim k^{\alpha} \quad (k \rightarrow 0)
\end{equation}
or, equivalently,
\begin{equation}
h(r)\sim-\frac{1}{r^{d+\alpha}} \quad (r \rightarrow +\infty),
\end{equation}
one can observe a number of different kinds of dependencies of the
asymptotic number variance $\sigma^2$ on the window radius $R$ for
$R \rightarrow \infty$: \cite{hyperuniform, chase, chase2}
\begin{equation}
\label{eq_S0} \sigma^2(R) \sim \left \{
\begin{array}{c@{\hspace{0.3cm}}c@{\hspace{0.3cm}}c}
R^{d-1}\ln R, & \alpha = 1, &
\\ R^{d-\alpha}, & \alpha<1, &
\\ R^{d-1}, & \alpha>1, & \end{array} \right .
\end{equation}
Note that in all cases, the number variance of a hyperuniform
point pattern grows more slowly than $R^d$.

\section{IV. EQUILIBRIUM PHASE BEHAVIOR OF TRUNCATED TETRAHEDRA}

\begin{figure}[!ht]
\begin{center}
$\begin{array}{c}
\includegraphics[width=0.35\textwidth]{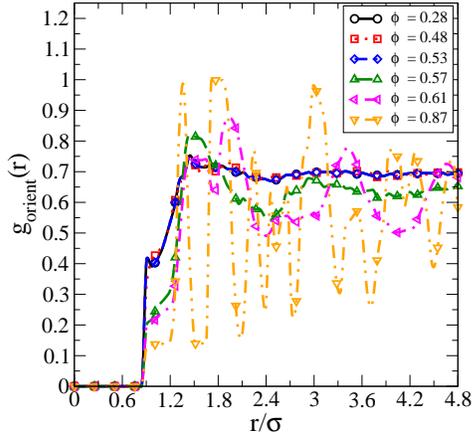}
\end{array}$
\end{center}
\caption{(Color online) Orientational pair correlation function $g_{orient}(r)$ as a function
of the dimensionless distance $r/\sigma$ (where $\sigma=\textnormal{v}^{1/3}$) for equilibrated systems of hard truncated tetrahedra at different packing fractions $\phi = 0.28, 0.48, 0.53, 0.57, 0.61$ and $0.87$. Below $\phi\approx0.53$ there is no long-range orientational correlation in the system, while around $\phi=0.57$ long-range orientational correlation arises.}
\label{fig_orient_corr}
\end{figure}

To explore the possible phases arising in the system of hard
truncated tetrahedra, we first carry out ASC simulations and
quantify the orientational order in the system. Specifically, we
compute the orientational pair correlation function [as defined in
eq \ref{eq_orient_corr}] for equilibrated systems with $N=686$
particles at different densities. As shown in Figure \ref{fig_orient_corr}, below $\phi\approx0.53$, there is no
long-range orientational correlation in the system, while around
$\phi=0.57$ long-range orientational correlation begins to emerge.
Together with the centroidal pair correlation function $g_{2}$
characterizing the translational order reported previously,
\cite{jiao2011communication} we find that the orientational order
and translational order arise almost simultaneously during a
possible first-order liquid-solid phase transition. These results strongly
suggest that the truncated tetrahedron system possesses neither
stable nematic nor rotator phases.

\subsection{A. First-Order Liquid-Solid Phase Transition of Truncated Tetrahedra}

\begin{figure}[!ht]
\begin{center}
$\begin{array}{c}
\includegraphics[width=0.35\textwidth]{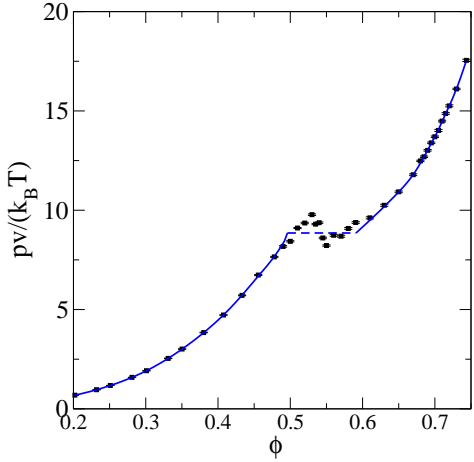}
\end{array}$
\end{center}
\caption{(Color online) Reduced pressure $p\textnormal{v}/(k_{B}T)$ as a function of
packing fraction $\phi$ for the
liquid branch (below $\phi\approx0.49\pm0.01$), liquid-solid phase
transition region (between $\phi\approx0.49\pm0.01$ and
$\phi\approx0.59\pm0.01$) and the crystal branch (above
$\phi\approx0.59\pm0.01$) of hard truncated tetrahedra, where v is the volume of a truncated tetrahedron.
The square dots are the actual simulation data and the solid curves are the fits of the data
for the liquid and crystal branches, respectively. The horizontal dash line denotes the coexistence
region between the liquid and the CT crystal, which is
determined by subsequent free energy calculations.}
\label{fig_trun_sl_pressure}
\end{figure}

We employ pressure and free energy calculations to determine
precisely the freezing- and melting-point packing fractions for hard truncated
tetrahedra. Specifically, we calculate the reduced pressure, i.e, $p\textnormal{v}/(k_{B}T)$
vs. $\phi$ for the liquid branch, liquid-solid phase
transition region, and the crystal branch, where v is the volume of a truncated tetrahedron. A periodic simulation
box containing $N=686$ hard truncated tetrahedra is employed. A
system at each density is initially generated by dilating
the fundamental cell of the densest crystal of the particles,
which is then equilibrated. At each density, at least 500000 MC
trial moves per particle and 5000 trial volume-preserving shear deformations of the fundamental
cell are applied to equilibrate
the system. Then the pressure of the equilibrated system is
collected. As shown in Figure \ref{fig_trun_sl_pressure}, the reduced
pressure $p\textnormal{v}/(\rho k_{B}T)$ increases smoothly with $\phi$ along the
equilibrium liquid branch up to a density $\phi\approx0.49\pm0.01$. Within the
density range between $\phi\approx0.49\pm0.01$ and $\phi\approx0.59\pm0.01$, the
trend of the reduced pressure exhibits discontinuities, suggesting
that metastable states exist in this region. (These coexistence densities
will be determined more precisely by free energy calculations below.)
Beyond $\phi\approx0.59\pm0.01$, the trend becomes smooth again,
suggesting that the system enters an equilibrium solid branch.

\begin{figure}[!ht]
\begin{center}
$\begin{array}{c}
\includegraphics[width=0.35\textwidth]{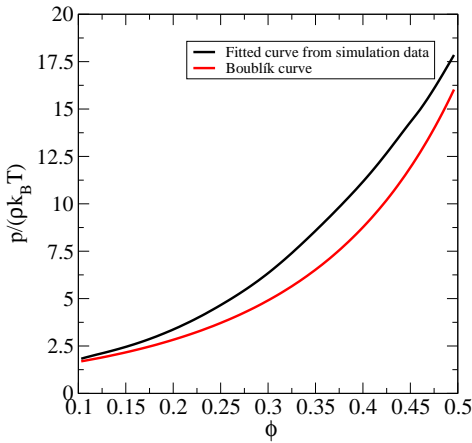}
\end{array}$
\end{center}
\caption{(Color online) $Z=p/(\rho k_{B}T)$ as
a function of packing fraction $\phi$  for the liquid branch by Boubl\'{\i}k's
expression and simulation data, respectively.
Note that the two curves diverge appreciably.}
\label{fig_eos}
\end{figure}

For the liquid branch, we compare the fit of our data for the dimensionless pressure $Z=p/(\rho k_{B}T)$
to the Boubl\'{\i}k's analytical approximation for the EOS \cite{bb_eos} for convex hard particles, which is given by
\begin{equation}
Z=\frac{1}{1-\phi}+\frac{3A\phi}{(1-\phi)^2}+\frac{3A^2\phi^2-A(6A-5)\phi^3}{(1-\phi)^3} \label{eq_eos}
\end{equation}
where $A=S\bar{R}/3\textnormal{v}$ is the nonsphericity parameter ($S$, $\bar{R}$, $\textnormal{v}$ are the surface area, radius of mean curvature, volume of a single particle, respectively).
As shown in Figure \ref{fig_eos}, the Boubl\'{\i}k's EOS appreciably underestimates the simulation data, which indicates the need for an improved analytical approximation for the EOS.

\begin{figure}[!ht]
\begin{center}
$\begin{array}{c}
\includegraphics[width=0.35\textwidth]{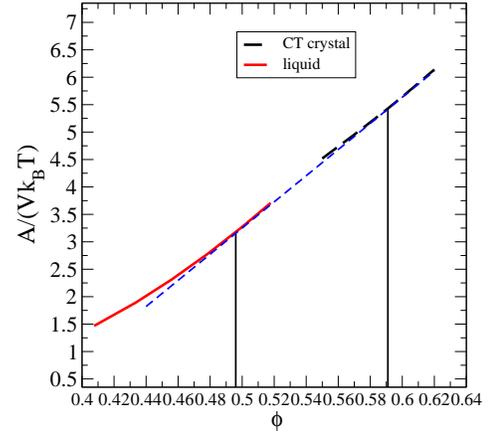}
\end{array}$
\end{center}
\caption{(Color online) Reduced Helmholtz free energy per unit volume $A/(Vk_{B}T)$ as a function of packing fraction
$\phi$ in the vicinity of the liquid-CT crystal coexistence region of hard truncated tetrahedra. The freezing- and melting-point packing fractions of hard truncated tetrahedra are
estimated to be $\phi=0.496\pm0.006$ and $\phi=0.591\pm0.005$,
respectively.} \label{fig_trun_sl_energy}
\end{figure}

To precisely locate the freezing- and melting-point packing fractions associated
with this first-order transition, we calculate the free energies for
the liquid and CT crystal phase at different densities. For the
liquid phase, a periodic simulation box containing $N=686$ hard
truncated tetrahedra is employed. At each density, at least 500000
MC trial moves per particle and 5000 trial volume-preserving shear deformations of
the fundamental cell are applied to equilibrate
the system. For the CT crystal phase, to eliminate finite-size
effects, we use four different system sizes ($N=432$, $N=1024$, $N=1458$, and
$N=2000$) and extrapolate the corresponding free energies
to obtain the infinite-size limit.\cite{polson2000finite} To
suppress particle collisions, $\lambda_{max}$ in eq
\ref{eq_energy_solid} is chosen to be 2000. For each system, 40000
MC cycles (i.e., a sequential trial move of each particle) are
used to equilibrate the system and another 40000 cycles are used
to compute the ensemble-averaged value of $U(\lambda)$ at each
integration point in eq \ref{eq_energy_solid}.

After the Helmholtz free energy per unit volume $A/(Nk_{B}T)$ as a function of $\phi$ for different
phases are obtained, we employ a common tangent construction, as
shown in Figure \ref{fig_trun_sl_energy}, to find precisely the
coexistence densities. \cite{ni2012phase} We find that the
freezing- and melting-point packing fractions are respectively given by $\phi=0.496\pm0.006$
and $\phi=0.591\pm0.005$.

\subsection{B. First-Order Solid-Solid Phase Transition of Truncated Tetrahedra}

Jiao and Torquato suggested that there is a solid-solid phase
transition between the CT crystal and the densest known crystal at
high densities by monitoring structural changes upon decompression from the densest crystal. \cite{jiao2011communication} However, the nature
of this transition and the exact transition densities were not
determined in their study. Here we employ pressure and free-energy
calculations to investigate this putative solid-solid phase transition.

\begin{figure}[!ht]
\begin{center}
$\begin{array}{c}
\includegraphics[width=0.35\textwidth]{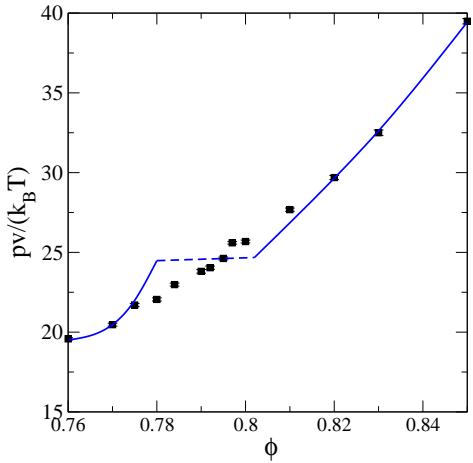}
\end{array}$
\end{center}
\caption{(Color online) Reduced pressure $p\textnormal{v}/(k_{B}T)$ as a function of packing fraction $\phi$ at high
densities for truncated tetrahedra, where v is the volume of a truncated tetrahedron. The square dots
are the actual simulation data and the solid curves are fits of the data for the CT and densest known crystal phases, respectively. The horizontal dash line denotes the coexistence region between these two crystals, which is determined by subsequent free energy calculations.
Note that approximately between $\phi\approx0.775\pm0.005$ and
$\phi\approx0.800\pm0.002$ the trend for simulation data exhibits
discontinuities, suggesting the occurrence of a solid-solid phase
transition.} \label{fig_trun_ss_pressure}
\end{figure}

Specifically, we calculate the reduced pressure $p\textnormal{v}/(k_{B}T)$ as a function of $\phi$ for
$\phi \in [0.76, 0.85]$. A periodic simulation box
containing $N=686$ hard truncated tetrahedra is employed. At each
density, at least 500000 MC trial moves per particle and 5000
trial volume-preserving shear deformations of the fundamental cell
are applied to equilibrate the system. The pressure of the
equilibrated system is then collected. As shown in Figure \ref{fig_trun_ss_pressure}, between $\phi\approx0.775\pm0.005$ and
$\phi\approx0.800\pm0.002$, the trend of the reduced pressure $p\textnormal{v}/(k_{B}T)$
versus $\phi$ exhibits a weak discontinuity, suggesting a
possible first-order solid-solid phase transition, which is
consistent with the qualitative study of Jiao and Torquato \cite{jiao2011communication}
in which structural changes upon decompression were monitored.

\begin{figure}[!ht]
\begin{center}
$\begin{array}{c}
\includegraphics[width=0.35\textwidth]{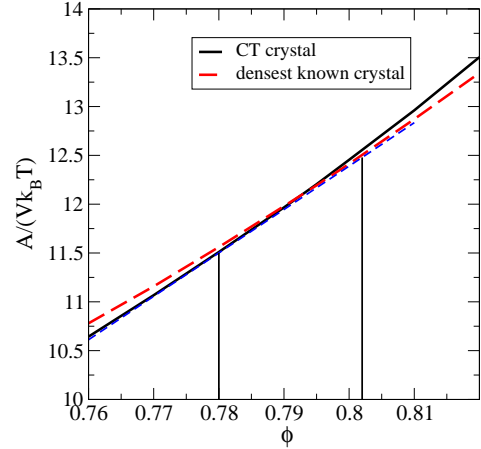}
\end{array}$
\end{center}
\caption{(Color online) Reduced Helmholtz free energy per unit volume $A/(Vk_{B}T)$ as a function of packing fraction
$\phi$ for CT crystal phase and densest known crystal phase of hard truncated tetrahedra. The coexistence densities of
CT crystal and densest known crystal are estimated to be $\phi=0.780\pm0.002$ and $\phi=0.802\pm0.003$.}
\label{fig_trun_ss_energy}
\end{figure}

\begin{figure}[!ht]
\begin{center}
$\begin{array}{c}
\includegraphics[width=0.35\textwidth]{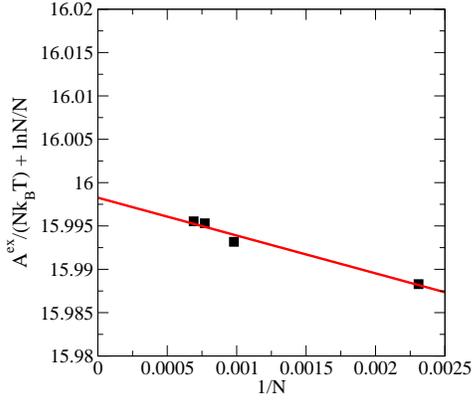}
\end{array}$
\end{center}
\caption{(Color online) The quantity $A^{ex}/(Nk_{B}T)+\ln N/N$ as a function of $1/N$ for a
system of truncated tetrahedra in a CT packing crystal at a
packing fraction of 0.78, where $A^{ex}\equiv A-A_{id}$ is the excess
free energy ($A_{id}$ is the ideal gas free energy). The intercept of the ordinate obtained from the linear extrapolation yields the infinite-volume limit of the excess free energy.}
\label{fig_finite_size}
\end{figure}

To precisely locate the coexistence densities associated with this
transition, we calculate the free energy for the CT crystal and
densest known crystal at different densities, as shown in Figure \ref{fig_trun_ss_energy}. As in the case
of liquid-solid transition, we correct for finite-size effects
in the values of the free energy at some fixed density by using
different system sizes (i.e., $N=432$, $N=1024$,
$N=1296$, and $N=1458$) and extrapolation (see Figure \ref{fig_finite_size}). The parameter $\lambda_{max}$ in eq
\ref{eq_energy_solid} is chosen to be 12000 to suppress particle
collisions. For each system, 40000 MC cycles are used to
equilibrate the system and another 40000 cycles are used to
collect the ensemble averaged value of $U(\lambda)$ at each
integration point. A common tangent construction is then employed
to identify the coexistence densities for the first-order CT to
densest-known crystal transition, i.e., $\phi \in [0.780\pm0.002,
~0.802\pm0.003]$.

\section{V. MAXIMALLY RANDOM JAMMED PACKINGS OF TRUNCATED TETRAHEDRA}
As indicated in section I, a hard-particle system falls out of
equilibrium when compressed sufficiently fast to
a disordered jammed packing state. The largest possible
compression rate consistent with jamming will result in the maximally random jammed
packing state. In this section, we generate the MRJ
packings of truncated tetrahedra using ASC simulations with a
sufficiently large compression rate and then study their structural characteristics.

\subsection{A. Generation of MRJ Packings via Fast Compression Using the ASC Scheme}
Starting from an unjammed initial packing configuration, the
particles are randomly displaced and rotated sequentially. If a
trial move (e.g., random displacement or rotation of a particle)
causes overlap between a pair of particles, it is rejected;
otherwise, the trial move is accepted and a new packing
configuration is obtained. After a prescribed number of particle
trial moves, small random deformations and compressions/dilations
of the simulation box are applied such that the system is on
average compressed. The compression rate $\Gamma$ is defined as
the inverse of the number of particle trial moves per particle per
simulation-box trial move. For large $\Gamma$, the system cannot
be sufficiently equilibrated after each compression and will
eventually jam with a disordered configuration at a lower density
than that of the corresponding maximally dense crystalline
packing. \cite{mrj3}

Two types of unjammed packings are used as initial configurations:
dilute equilibrium hard polyhedron fluids with $\phi < 0.1$ and
packings derived from MRJ hard-sphere packings. In the later case,
the largest possible polyhedron with random orientation is inscribed
into each sphere, which is employed to maximize both translational and
orientational disorder in the initial packings. Initial
configurations of both types are quickly compressed ($\Gamma \in
[0.01, 0.1]$) to maximize disorder until the average interparticle
gap is $\sim0.1$ of the circumradius of the polyhedra. Then a much
slower compression ($\Gamma \in [0.0002, 0.001]$) is used to
allow true contact network to be established which induces
jamming. The final packings are verified to be strictly jammed by
shrinking the particles by a small amount ($<0.01$ circumradius )
and ``equilibrating'' the system with a deformable boundary. If there is
no increase of the interparticle gaps (decreasing pressure) for a sufficiently long period of time ($>50~000$ MC
moves per particle), the original packing is considered to be
jammed. \cite{mrj3} Translational and orientational order
are explicitly quantified by evaluating corresponding correlation
functions, which then enables us to find those configurations with
a minimal degree of order among a representative set of
configurations. This analysis leads to reasonably close
approximations to the MRJ states. \cite{mrj3}

\begin{figure}[!ht]
\begin{center}
$\begin{array}{c}
\includegraphics[width=0.40\textwidth]{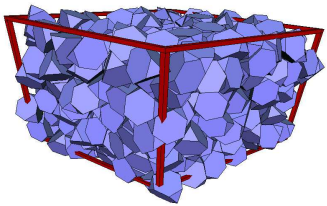}
\end{array}$
\end{center}
\caption{(Color online) Representative MRJ packing of truncated tetrahedra with $N=500$ particles.}
\label{fig_MRJ}
\end{figure}

\begin{figure}[!ht]
\begin{center}
$\begin{array}{c}
\includegraphics[width=0.35\textwidth]{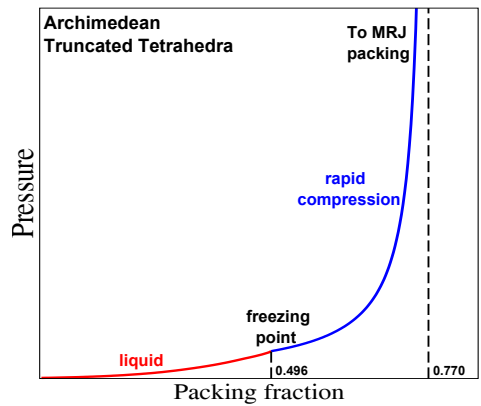}
\end{array}$
\end{center}
\caption{(Color online) Schematic plot of the pressure of the hard truncated tetrahedra system as a function of packing fraction along the liquid branch and then along a metastable extension of the liquid branch that ends at
the MRJ state. Note that the pressure diverges at the jamming point $\phi\approx0.770$, which is substantially larger than for spheres ($\phi_{MRJ} \approx 0.64$); see Figure \ref{fig_sphere_phase}.}
\label{fig_mrj_pressure}
\end{figure}

For truncated tetrahedra, jammed final packings with similar
$\phi$ and structural characteristics can be obtained from both
types of initial configurations. Although larger $\Gamma$ than
employed here can lead to final packings with even lower $\phi$
and a higher degree of disorder, such packings are generally not
jammed, i.e., they are ``melt'' upon small shrinkage and
equilibration. We have used the largest possible initial
compression rates ($\Gamma \in [0.01, 0.1]$) that lead to jammed
packings. The packings studied here contain $N=500$ particles.
We find that the average packing fraction of the MRJ state of truncated tetrahedra
is $\phi_{MRJ} = 0.770 \pm 0.001$. A representative
MRJ packing is shown in Figure \ref{fig_MRJ}.
Figure \ref{fig_mrj_pressure} schematically depicts the pressure of the hard truncated tetrahedra system
as a function of packing fraction. Note that the pressure diverges at the
jamming point $\phi\approx0.770$, which is substantially larger
than for spheres ($\phi_{MRJ} \approx 0.64$);
see Figure \ref{fig_sphere_phase}.

\subsection{B. Pair Correlations in MRJ Packings of Truncated Tetrahedra}

\begin{figure}[!ht]
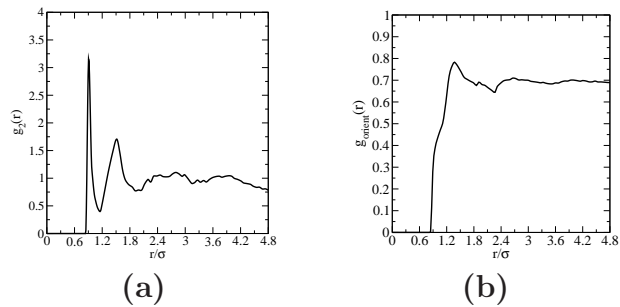

\begin{center}
$\begin{array}{c@{\hspace{1.0cm}}c}\\
\includegraphics[width=0.2\textwidth]{fig13a.eps} &
\includegraphics[width=0.2\textwidth]{fig13b.eps} \\
\mbox{\bf (a)} & \mbox{\bf (b)}
\end{array}$
\end{center}
\caption{(Color Online) Pair correlation functions versus the dimensionless distance $r/\sigma$
(where $\sigma=\textnormal{v}^{1/3}$) for MRJ packings of truncated tetrahedra obtained by averaging over 5 configurations: (a) centroidal pair correlation function $g_2(r)$ and (b) orientational pair correlation function $g_{orient}(r)$.}
\label{fig_g2}
\end{figure}

We find the packing fraction of the MRJ states of truncated
tetrahedra is $\phi = 0.770 \pm 0.001$. A representative packing
configuration is shown in Figure \ref{fig_MRJ}. The panels of Figure \ref{fig_g2} show the pair correlation
function $g_2(r)$ associated with the particle centroids and the
orientational pair correlation function $g_{orient}(r)$ obtained by
averaging over 5 configurations. It can be
seen that $g_2(r)$ possesses several prominent oscillations.
However, the magnitude of these oscillations are much smaller than
that associated with MRJ packings of hard spheres, or other
polyhedra with small asphericity values (e.g., icosahedra). This
is because in MRJ sphere packings, the pair distances between
contacting neighbors are exactly equal to the diameter of the
spheres. However, for nonspherical particles, the pair distances
between contacting neighbors in the associated MRJ packings can
vary from the diameter of their insphere to that of their
circumsphere, and thus, causing large fluctuations of pair
distances between the particle centroids. This further diminishes
the magnitude of the oscillations in the associated $g_2(r)$, and
thus the translational order in the system. The orientational
correlation function $g_{orient}(r)$ also suggests that the packing is orientationally disordered.

\begin{figure}[!ht]
\begin{center}
$\begin{array}{c}
\includegraphics[width=0.35\textwidth]{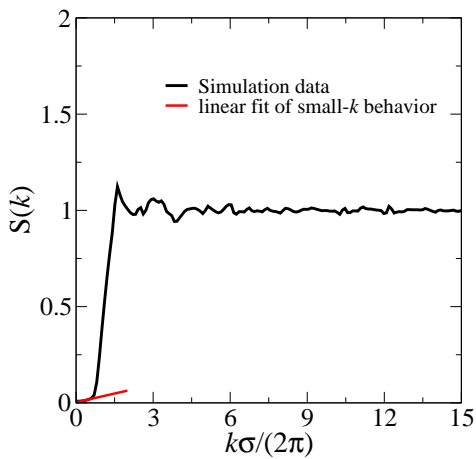}
\end{array}$
\end{center}
\caption{(Color online) Structure factor $S(k)$ as a function of the dimensionless wavenumber $k\sigma/(2\pi)$ obtained by averaging over 5 configurations of MRJ packings of truncated tetrahedra, where $\sigma=\textnormal{v}^{1/3}$. The red line is a linear fit to the data for small wavenumbers.}
\label{fig_Sk}
\end{figure}

Figure \ref{fig_Sk} shows the structure factor $S(k)$ versus wavenumber $k$ as obtained by averaging over 5 configurations of MRJ packings of truncated tetrahedra. Importantly, we find that
$S(k)\rightarrow 0$ as $k \rightarrow 0$, which means they are
hyperuniform with quasi-long-range pair correlations (see Sec. III.C). We employ a linear polynomial to approximate
the small-$k$ behavior of $S(k)$, i.e., $S(k)=a_0 + a_1 k\sigma/(2\pi)$, and
use it to fit computed $S(k)$ and find that
$a_0 \approx 0$ ($<10^{-5}$). Moreover, the slope is $a_1 = 0.030$, which is
significantly larger than $a_0$. These observations indicate that
the MRJ truncated tetrahedron packings possess hyperuniform quasi-long-range
(QLR) pair correlations in which $h(r)$ decays asymptotically with scaling
$-r^{-4}$, consistent with our previous studies of MRJ packings of
the non-tiling Platonic solids. This also provide further evidence
that the hyperuniform QLR correlations is a universal signature of
MRJ packings of hard convex particle of general shape. \cite{chase_prl}
Note that although we are employing a relatively small system with $N=500$ particles,
based on previous research \cite{mrj3} we expect that the extrapolated small-$k$ behavior of $S(k)$
obtained from such systems should agree with extrapolations from large systems. Therefore, our conclusions that MRJ packings of truncated tetrahedra are hyperuniform, based on the relatively small systems
that we have studied, should be firm.

It is interesting to compare the slope $a_1$ of $S(k)$ for small $k$ of truncated
tetrahedra to those of other polyhedra that we have previously studied. For icosahedra,
dodecahedra, octahedra and tetrahedra, the values of $a_1$ are
respectively $0.015$, $0.023$, $0.029$ and $0.21$. It is apparent
that as the polyhedral shape deviates more from that of a sphere,
the value of the slope $a_1$ increases, i.e., the degree of
hyperuniformity decreases. This is because larger asphericities
induce larger local number density fluctuations at fixed long
wavelengths (i.e., small $k$ values) due to the QLR correlations.

We would like to note that for nonspherical particles, the more appropriate measure
of hyperuniformity is based on local volume
fraction fluctuations (rather than density fluctuations) and the small-wavenumber behavior of the
associated spectrum density. \cite{chase_prl} Similar conclusions on hyperuniform QLR correlations in the system
will be obtained if the more general procedure is used, but
with this procedure, the degree of hyperuniformity would decrease
with decreasing packing fraction (instead of with increasing
asphericity). This is because denser packings possess more
homogeneous void spaces, which lead to smaller local volume
fraction fluctuations. \cite{chase_prl}

Table \ref{tab_mrj} lists the values of the MRJ packing fraction
$\phi_{MRJ}$ and the packing fraction of the densest known packing
$\phi_{max}$ for different particle shapes. \cite{sal_2009nature,
torquato2009dense, mrj2, mrj3, superball_dense,
jiao2011communication, ellipsoid_1, ellipsoid_2, sphere_1,
sphere_2, sphere_3, tetra_ref} It is interesting to note that the
ratio $\beta = \phi_{MRJ}/\phi_{max}$ provides a measure of the
extent to which the system is a good glass former. Specifically,
when $\beta$ is close to unity, high-density disordered packings
are entropically favored, i.e., it is more difficult to produce
the crystal phase and the associated maximally dense packing via
simulation or experimental protocols. Indeed, for spheroids with
$\beta= 0.95$, the densest-known packing was extremely difficult
to obtain via simulations, which requires very slow compression of
a small system with a few particles in a fundamental cell with a
specific shape. \cite{ellipsoid_1} It has recently been shown that
disordered binary sphere packings can attain MRJ packing densities
that are nearly as dense as the densest known packings and
therefore are good glass formers. \cite{binary_ref} Finally, consistent with the results for MRJ packings of non-tiling Platonic solids, \cite{mrj3} we find that the rattler-free jammed ``backbones'' \cite{rattler} MRJ packings of truncated tetrahedra are also isostatic (the total number of constraints, related to the different types of interparticle contacts, equals the total number of degrees of freedom in the packing). Specifically, each particle in MRJ packings of truncated tetrahedra has 12.01 contacts on average (the number of face-to-face, edge-to-face, vertex-to-face, and edge-to-edge contact is determined to be $2.79 \pm 0.02$, $0.32 \pm 0.02$, $0.92 \pm 0.02$ and $2.07 \pm 0.02$, respectively), which is consistent with isostaticity.

\begin{table}
\caption{Relationship between the MRJ Packing Fraction $\phi_{MRJ}$ and the Densest Packing Fraction $\phi_{max}$ for Congruent Particles of Different Shapes. \cite{sal_2009nature, torquato2009dense, mrj2, mrj3, superball_dense, jiao2011communication, ellipsoid_1, ellipsoid_2, sphere_1, sphere_2, sphere_3, tetra_ref, tetra_ref2, tetra_ref3} (In the case of spheroids, $b/a$ is the aspect ratio and in the case of superballs, $q$ is the deformation parameter (where $q=1$ corresponds to a sphere). A ratio $\phi_{MRJ}/\phi_{max}$ close to unity indicates a system that is a good glass former, as discussed in the text.)}
\begin{center}
\begin{tabular}{>{\centering\arraybackslash}m{4cm}>{\centering\arraybackslash}m{1cm}>{\centering\arraybackslash}m{1cm}>{\centering\arraybackslash}m{2cm}} \\ \hline\hline
Particle shape & $\phi_{MRJ}$ & $\phi_{max}$ & $\phi_{MRJ}/\phi_{max}$ \\
\hline
Octahedron & 0.697 & 0.947 & 0.736 \\
\hline
Tetrahedron & 0.763 & 0.856 & 0.891 \\
\hline
Dodecahedron & 0.716 & 0.904 & 0.792 \\
\hline
Icosahedron & 0.707 & 0.836 & 0.846 \\
\hline
Truncated tetrahedron & 0.770 & 0.995 & 0.774 \\
\hline
Sphere & 0.637 & 0.741 & 0.86 \\
\hline
Prolate spheroid with $b/a=2$  & 0.70 & 0.77 & 0.91 \\
\hline
Prolate spheroid with $b/a=3/2$ & 0.71 & 0.75 & 0.95 \\
\hline
Oblate spheroid with $b/a=1/2$ & 0.70 & 0.77 & 0.91 \\
\hline
Oblate spheroid with $b/a=2/3$ & 0.71 & 0.75 & 0.95 \\
\hline
Superball with $q=0.8$ & 0.66 & 0.72 & 0.92 \\
\hline
Superball with $q=1.2$ & 0.68 & 0.77 & 0.88 \\
\hline
Superball with $q=2$ & 0.74 & 0.86 & 0.86 \\
\hline\hline
\end{tabular}
\end{center}
\label{tab_mrj}
\end{table}
\clearpage

\section{VI. CONCLUSIONS AND DISCUSSION}
In this paper, we have ascertained the equilibrium phase behavior
of truncated tetrahedra over the entire range of possible
densities, via ASC Monte Carlo simulations and free energy
calculations. We found that the system undergoes first-order
liquid-solid and solid-solid phase transitions as the density
increases, consistent with the finding in the qualitative
study by Jiao and Torquato. \cite{jiao2011communication} The
liquid phase coexists with the CT crystal phase within the density
range $\phi \in [0.496\pm0.006, ~0.591\pm0.005]$ and the CT phase
coexist with the densest-known crystal within the density range
$\phi \in [0.780 \pm 0.002, 0.802 \pm 0.003]$. We found no
evidence for any stable rotator or nematic phases. We also generated the
maximally random jammed (MRJ) packings of truncated tetrahedra,
which may be regarded to be the ending state of a metastable branch of the phase diagram
for truncated tetrahedra. Specifically, we systematically studied
the structural characteristics of the MRJ packings, including the
centroidal pair correlation function, structure factor and orientational pair
correlation function. We found that
such MRJ packings are hyperuniform with an average
packing fraction of $0.770 \pm 0.001$, which is considerably larger
than the corresponding value for identical spheres. We
have also shown that the ratio $\phi_{MRJ}/phi_{max}$ for a general nonspherical
particle shape provides a measure of its glass-forming ability.

The transition from the CT to the densest-known crystal phase
involves symmetry breaking as one would expect. Specifically, the
CT packing possess a higher degree of symmetry (rhombohedral) than
that of the densest known packing (triclinic). Similar
symmetry-breaking crystal-crystal phase transitions have also been
observed in systems of truncated cubes with a small degree of
truncation, \cite{gantapara2013phase} which possess a first-order
transition from the simple-cubic phase (with cubic symmetry) to
the one associated with the densest-known packing (with
rhombohedral symmetry). We conjecture that particles with
shapes close to the space-filling ones and that nearly fill all of
space (i.e., those obtained by chopping of the corners of
rhombohedron, cube, truncated octahedron, and certain prisms)
would probably undergo a first-order crystal-crystal transition
from a high-symmetry solid phase to a low-symmetry one. The
high-symmetry phase should be associated with the optimal configuration
of the corresponding space-filling shape, and the low-symmetry phase should be associated with the
densest-known packing of the actual particle that is nearly
space-filling. \cite{conway2006packing, jiao2011communication} Since
the low-symmetry phase should possess more free volume and thus is
more favorable entropically compared to the high-symmetry phase,
we expect that a solid-solid phase transition would occur at high
densities in such systems. However, we emphasize that such qualitative predictions about the class of particle shapes that possess crystal-crystal phase transitions has yet
to be identified and verified by rigorous free energy calculations.

These results also point to the great challenges in identifying
solid-solid phase transitions in hard-particle systems. In
particular, for truncated tetrahedra, if one had not known {\it a
prior} about the existence of both crystal structures (i.e., the
CT and densest known packings), \cite{conway2006packing,
jiao2011communication} one would never have tried to see if the
phases associated with these distinct crystals could coexist. For
example, if one only knew about the densest known packing but not
the CT packing, and simply decompressed the packing from the highest
density, one likely would have thought there was no other crystal
phase. Similarly, if one only knew about CT packing, then further
compression of the CT from the melting point would probably not
have alerted one to consider another crystal structure at very high
densities. This also calls into the question whether in
previous studies similar solid-sold phase transitions were missed
for other particle shapes because no systematic structural
probes were used to ascertain whether the crystal structure
changed symmetry. We believe that crystal structures with distinct
symmetries should be explicitly examined in order to correctly
identify possible solid-solid phase transitions.

It is also useful to compare the phase behavior of truncated
tetrahedra to that of other nonspherical shapes, especially the
nature of the disorder-order phase transition. Truncated
tetrahedra behave like regular octahedra \cite{ni2012phase} in that both
systems undergo a first-order transition from the isotropic liquid
to a crystal phase. It is clear that this transition, which is
entropy-driven, \cite{frenkel} is determined by the characteristics of the
particle shape. Both the truncated tetrahedron and regular octahedron have an
aspect ratio $\delta$ (defined to be the ratio between the longest
and shortest principal axis) of unity. They also possess
moderately sized values of a relative scaled exclusion volume $\tau
\equiv \textnormal{v}_{ex}/(8\textnormal{v})$, \cite{sal_perco1} i.e., $\tau = 1.236$ and
$1.330$ for truncated tetrahedra and octahedra, respectively. Upon
forming the crystal phase from the liquid, a sufficiently large
increase of free-volume entropy is achieved by particle alignment
and positional ordering, and thus, the crystal phase is favored. For
shapes close to that of a sphere, such as ellipsoids with small
$\delta$ and superballs with $\tau$ close to unity, the associated
systems possess a transition from isotropic to rotator phase. This
is because orientational ordering does not lead to significant
increase of free-volume entropy, and thus, is not favored. On the
other hand, as the aspect ratio of an ellipsoid increases beyond
a critical value (e.g., $\delta > 2$), the systems start to form a
nematic phase from liquid upon compression,
\cite{bautista2013further} due to a large gain of free-volume
entropy by particle alignment.

On the basis of the aforementioned observations and other previous research,
\cite{ni2012phase, gantapara2013phase, jiao2011communication, bautista2013further}
we predict the phase that breaks the symmetry of the liquid for systems of hard non-spherical
particles such as the regular icosahedra, dodecahedra, \cite{ref2012} and rectangular
parallelepiped (with side lengths $a$, $b$, and $c$, and $a\ne b\ne c$).
Specifically, we expect that when the relative scaled exclusion volume $\tau$ for particles
with flat faces is less than the order of $1.2$, it is highly possible that rotator phase
arises upon compression from isotropic liquid phase. The scaled exclusion volume of icosahedron is
given by $\textnormal{v}_{ex}/\textnormal{v}= [2+\frac{90\sqrt{3}\cos^{-1}(\sqrt{5}/3)}{\pi
(2+\sqrt{5})}] \approx 8.915$, \cite{sal_perco1} and the relative
scaled exclusion volume is $\tau=1.114$, which is very close to 1.
Therefore, we expect the icosahedron systems to possess an
isotropic-rotator phase transition. Similarly, the scaled
exclusion volume of dodecahedron is given by $\textnormal{v}_{ex}/\textnormal{v}=
[2+\frac{90\sqrt{25+10\sqrt{5}}\cos^{-1}(1/\sqrt{5})}{\pi
(15+7\sqrt{5})}] \approx 9.121$ and $\tau=1.140$.
\cite{sal_perco1} We see again that the later value is close to 1,
which suggests that the dodecahedron system is likely to possess a
rotator phase upon compression from isotropic liquid phase as
well. The scaled exclusion volume of rectangular
parallelepiped ($a\ne b\ne c$) is given by $\textnormal{v}_{ex}/\textnormal{v}=
[2+\frac{(ab+bc+ac)(a+b+c)}{abc}]$. Again, when the value of
$\tau$ is less than the order of $1.2$, we expect that there
exists a phase transition from the isotropic liquid to a rotator phase.
Rigorous free energy calculations will be able to verify such
predictions.





\begin{acknowledgement}

We are grateful to Ran Ni for very helpful discussions.
This work was supported in part by the National
Science Foundation under Grants No. DMR-0820341 and No.
DMS-1211087. This work was partially supported by a grant
from the Simons Foundation (Grant No. 231015 to S.T.).

\end{acknowledgement}


\begin{thebibliography}{78}
\bibitem{ref2012}
We note that such rotator phases were observed in Monte-Carlo simulations. \cite{Damasceno12} However, rigorous free energy calculations are still
required to verify that these phases are thermodynamically stable
rather than metastable.
\end{thebibliography}

\begin{tocentry}

\includegraphics[width=8.5cm]{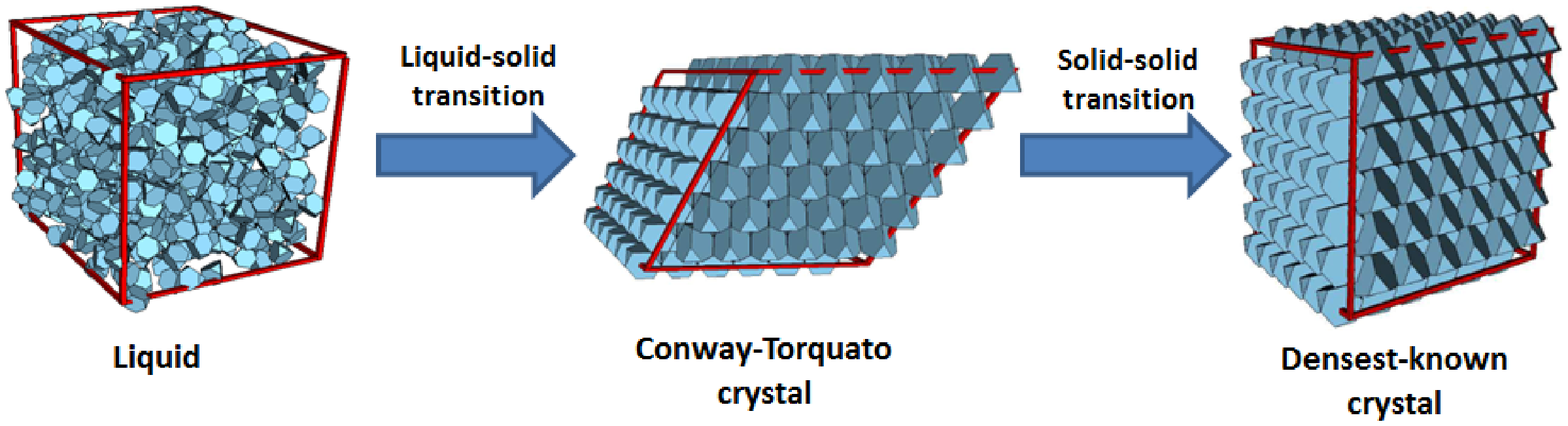}

\end{tocentry}

\end{document}